\pdfoutput=1
\documentclass[twocolumn, tighten]{myaastex62}
\usepackage{amsmath}
\shorttitle{Super-Earths in need for Extremly Big Rockets}
\shortauthors{Michael Hippke}
\begin{document}
\title{SUPER-EARTHS IN NEED FOR EXTREMLY BIG ROCKETS}

\author[0000-0002-0794-6339]{Michael Hippke}
\affiliation{Sonneberg Observatory, Sternwartestr. 32, 96515 Sonneberg, Germany}
\email{michael@hippke.org}

\begin{abstract}
Many rocky exoplanets are heavier and larger than Earth, so-called ``Super-Earths''. Some of these may be habitable, and a few may be inhabited by Super-Earthlings. Due to the higher surface gravity on these worlds, space-flight is much more challenging. We find that chemical rockets still allow for escape velocities on Super-Earths up to $10\times$ Earth mass. Much heavier rocky worlds, if they exist, will require using up most of the planet as chemical fuel for the (one) launch, a rather risky undertaking. We also briefly discuss launching rockets from water worlds, which requires Alien megastructures.\\
\end{abstract}

\section{Preface}
It should be noted that, while the subject of this paper is silly, the analysis actually does make sense. This paper, then, is a serious analysis of a ridiculous subject, which is of course the opposite of what is usual in astrophysics \citep[inspired by][]{Krugman2010}.

\section{Introduction}
Do we inhabit the best of all possible worlds (Leibniz 1710)? From a variety of habitable worlds that may exist, Earth might well turn out as one that is marginally habitable. Other, more habitable (``superhabitable'') worlds might exist \citep{2014AsBio..14...50H}. Planets more massive than Earth can have a higher surface gravity, which can hold a thicker atmosphere, and thus better shielding for life on the surface against harmful cosmic rays. Increased surface erosion and flatter topography could result in an ``archipelago planet'' of shallow oceans ideally suited for biodiversity. There is apparently no limit for habitability as a function of surface gravity as such \citep{2017arXiv171005605D}. Size limits arise from the transition between Terran and Neptunian worlds around $2\pm0.6\,R_{\oplus}$ \citep{2017ApJ...834...17C}. The largest rocky planets known so far are $\sim1.87\,R_{\oplus}$, $\sim9.7\,M_{\oplus}$  \citep[Kepler-20\,b,][]{2016AJ....152..160B}. When such planets are in the habitable zone, they may be inhabited by ``Super-Earthlings'' (SEALs). Can SEALs still use chemical rockets to leave their planet\footnote{What would SEALs look like? Perhaps like little green (wo)men with short thick legs needed for high surface gravity.}?

At our current technological level, spaceflight requires a rocket launch to provide the thrust needed to overcome Earth's force of gravity. Chemical rockets are powered by exothermic reactions of the propellant, such as hydrogen and oxygen. Other propulsion technologies with high specific impulses exist, such as nuclear thermal rockets \citep[e.g., NERVA,][]{1969JSpRo...6..565A}, but have been abandoned due to political issues. Rockets suffer from Tsiolkovsky's equation \citep{tsiolkovsky1903issledovanie}: if a rocket carries its own fuel, the ratio of total rocket mass versus final velocity is an exponential function, making high speeds (or heavy payloads) increasingly expensive \citep{1992CeMDA..53..227P}. While we hand-wave away many things in this paper, we do respect rocket science.

State-of-the-art technology such as the recently introduced Falcon Heavy \citep{Mann2018} has a rocket height of 70\,m, mass of 1421\,t, and delivers a payload of 16.8\,t to an Earth escape velocity, so that the payload fraction is $\sim1\,$\%. We will now explore how much rocket is needed for planets with higher surface gravity.


\section{Method}
The achievable maximum velocity change of a chemical rocket is

\begin{equation}
    \Delta v = v_\text{ex} \ln \frac{m_0}{m_f}
\end{equation}

where $m_{0}$ is the initial total mass (including fuel), $m_{f}$ is the final total mass without fuel (the dry mass), and $v_{\rm ex}$ is the exhaust velocity. We can substitute $v_{\rm ex}=g_{\rm 0}\,I_{\rm sp}$ where $g_{\rm 0}=G\,M_{\oplus}/R^2_{\oplus}\sim9.81\,$m\,s$^{-1}$ is the standard gravity and $I_{\rm sp}$ is the specific impulse (total impulse per unit of propellant), typically $\sim350\dots450\,$s for hydrogen/oxygen (for references, read Wikipedia).

To leave Earth's gravitational influence, a rocket needs to achieve at minimum the escape velocity

\begin{equation}
    v_{\rm esc} = \sqrt{\frac{2GM_{\oplus}}{R_{\oplus}}}\sim11.2\,{\rm km\,s}^{-1}
\end{equation}

for Earth, and $v_{\rm esc}\sim27.1\,$km\,s$^{-1}$ for a $10\, M_{\oplus}$, $1.7\,R_{\oplus}$ Super-Earth similar to Kepler-20\,b.

\section{Results}
We consider a single-stage rocket with $I_{\rm sp}=350\,$s and wish to achieve $\Delta v > v_{\rm esc}$. The mass ratio of the vehicle becomes

\begin{equation}
    \frac{m_0}{m_f} > {\rm exp} \left( \frac{v_{\rm esc}}{v_\text{ex}} \right).
\end{equation}

which evaluates to a mass ratio of $\sim26$ on Earth, and $\sim2{,}700$ on Kepler-20\,b. Consequently, a single-stage rocket on Kepler-20\,b must burn $104\times$ as much fuel for the same payload ($\sim2{,}700$\,t of fuel for each t of payload). This example neglects the weight of the rocket structure itself, and is therefore a never achievable lower limit. In reality, rockets are multistage, and have typical mass ratios (to Earth escape velocity) of $50\dots150$. For example, the Saturn~V had a total weight of 3{,}050\,t for a lunar payload of 45\,t, so that the ratio is 68. The Falcon Heavy has a total weight of 1{,}400\,t and a payload of 16.8\,t, so that the ratio is 83.

\begin{figure}
\includegraphics[width=\linewidth]{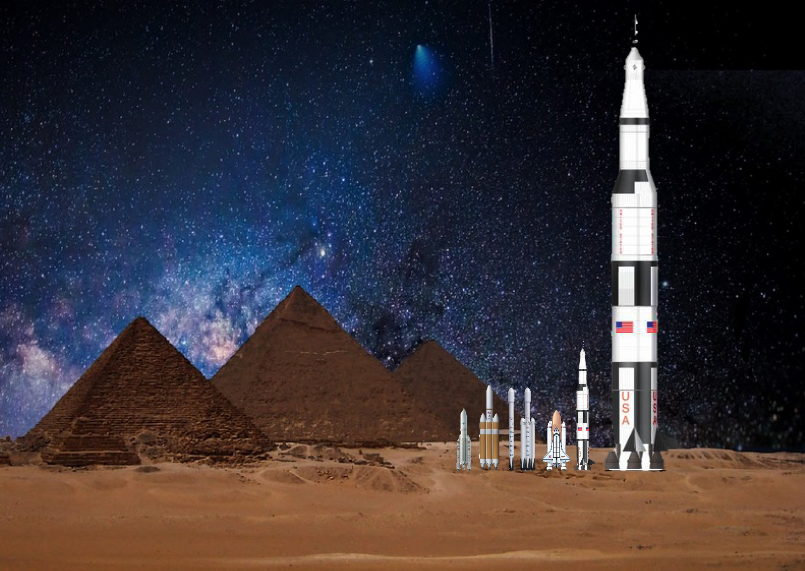}
\caption{\label{fig:rocket}Size comparison (left to right): Great Pyramids of Giza, Ariane~5, Delta Heavy, Falcon~9, Falcon Heavy, Space Shuttle, Saturn~V and the Extremely Big F*ing Rocket for a launch on Kepler-20\,b.}
\end{figure}

For a mass ratio of 83, the minimum rocket (1\,t to $v_{\rm esc}$) would carry $9{,}000\,$t of fuel on Kepler-20\,b, which is only $3\times$ larger than a Saturn~V (which lifted 45\,t). To lift a more useful payload of 6.2\,t as required for the James Webb Space Telescope on Kepler-20\,b, the fuel mass would increase to $55{,}000$\,t, about the mass of the largest ocean battleships. We show such a rocket to-scale in Figure~\ref{fig:rocket}. For a classical Apollo moon mission (45\,t), the rocket would need to be considerably larger, $\sim400{,}000$\,t. We are not sure how ridiculous such a rocket is, because it is still less heavy than the Pyramid of Cheops, although not by much.

\section{Discussion}

\subsection{Launching from a mountain top}
Rockets work better in space than in an atmosphere. How about launching the rocket from a high mountain? At first glance, this is a great idea, because the rocket thrust is given by

\begin{equation}
F = \dot{m}\,v_{ex} + A_e(P_1 - P_2)
\end{equation}

where $\dot{m}$ is the mass flow rate, $A_e$ is the cross-sectional area of the exhaust jet, $P_1$ is the static pressure inside the engine, and $P_2$ is the atmospheric pressure. The exhaust velocity is maximized for zero atmospheric pressure, i.e. in vacuum. Unfortunately, the effect is not very large in practice. For the Space Shuttle's main engine, the difference between sea level and vacuum is $\sim25\,$\% \citep{Rocketdyne}. Atmospheric pressure below 0.4\,bar (Earth altitude $6{,}000$\,m) is not survivable long term for humans, and presumably neither for SEALs. Then, the effect is $\sim15\,$\%. Such low pressures are reached in lower heights on Super-Earths, because the gravity pulls the air down. Strongly.

\subsection{But there is no mountain top...}
Another effect which is to the SEALs disadvantage is that the bigger something is, the less it can deviate from being smooth. Tall mountains will crush under their own weight \citep[the ``potato radius'' is $\sim238\,$km,][]{2015arXiv151104297C}. Therefore, we expect more massive planets to have smaller mountains. This will be detectable through transit observations in future telescopes \citep{2018MNRAS.tmp..142M}.

Indeed, the largest mountains in our solar system are on less massive bodies\footnote{\url{https://en.wikipedia.org/wiki/List_of_tallest_mountains_in_the_Solar_System}}. We recommend that the SEALs use shovels to make a gigantic mountain, exceeding the atmosphere, and launch their rocket from the vacuum on top. We encourage further research in this rather under-explored field.

\subsection{Launching rockets from water-worlds}
Many habitable (and presumably inhabited) planets might be waterworlds \citep{2017MNRAS.468.2803S}, and intelligent life in water and sub-surface is plausible \citep{2017arXiv171109908L}. How would Nautical Super Earthlings (Navy-SEALs) launch their rockets? This is actually less absurd than most other things in this paper, but harder than the reader might think. 

An elegant method would be to build an Alien ``megastructure'', as postulated by \citet{2016ApJ...816...17W}. To launch the rocket, a large floating structure can be used. Turtles do not exist in such sizes (citation needed), but nautical floats can be built in turtle shapes. The rocket would be placed on the turtle's shield (out of the water), dried with towels, and then launched towards the heavens, while other (living) turtles spray water towards the launch pad turtle to cool down the hot exhaust fumes.\footnote{Indeed, such an object was recently announced \citep[``Tabby-Star''][]{2016MNRAS.457.3988B} where Aliens are building a megastructure \citep[see][]{2016ApJ...829L...3W}. Ongoing rocket launches eject large quantities of water into space, forming a non-dusty orbital cloud around the star, perhaps leading to century-long dimming \citep{2016ApJ...822L..34S,2016ApJ...830L..39M,2016arXiv160502760L,2016ApJ...825...73H,2017ApJ...837...85H,2018ApJ...854L..11H}.}

These minor aquatic launch complications make the theory of oceanic rocket launches appear at first quite alien; presumably land-based launches seem equally human to alien rocket scientists.

\begin{figure}
\includegraphics[width=\linewidth]{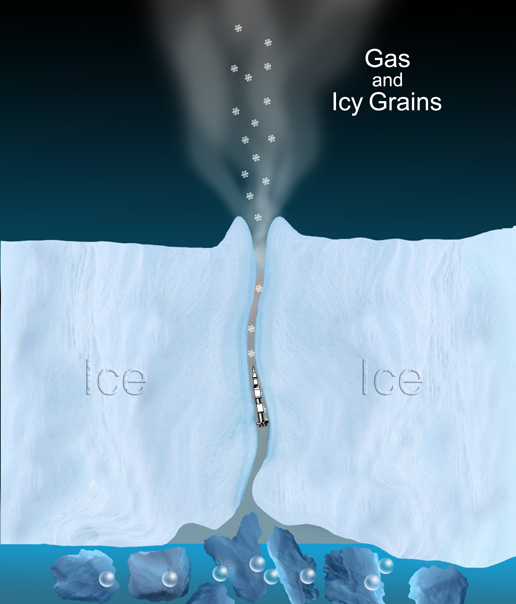}
\caption{\label{fig:ice}Extremely Big F*ing Rocket blasting through a plume of an Enceladus-like icy Super-Earth.}
\end{figure}

\subsection{Launching rockets on worlds with an icy crust}
Subsurface liquid water oceans exist below the frozen surfaces of Enceladus and Europa, and it appears plausible that such worlds are habitable. How would ice nautical Super-Earthlings (iNavy SEALS) launch their chemical rockets? They need an icebreaker.

One method which works\footnote{\url{https://www.youtube.com/watch?v=1aPvGGvnAGQ}} (sometimes not so well\footnote{See video of failed underwater submarine rocket launch, \url{https://www.popularmechanics.com/military/weapons/a25176/launching-missile-from-submarine/}}) is to use classical explosives to flash-vaporize water into steam. The pressure of the expanding gas drives the missile upwards in a tube. This works well for comparably small ICBMs launched from submerged submarines, but these have no issues with ice coverage. As a minor annoyance, ice crusts on Europa and Enceladus are tens of kilometers thick. A series of fusion bombs could be used to blast through the ice and then, quickly, lift the massive rocket completely out of the water. This is highly non-trivial, because of the vacuum on the outside, which does not allow for a liquid phase of the water. The expanding vaporized fountain would re-freeze quickly, leaving little time for the journey.

Fortunately for the fellow iNavy-SEALS who must stay behind, water (H$_2$O) cannot become radioactive itself, and radioactive particles are mostly not soluble in water. Therefore, they can be filtered out after the launch. Typical fallout particles sink to the sea floor in a few days, and in the meantime, drinking water can be drawn from near the top of the pool. Among the authors who have not pointed this out is \citet{ANDP:ANDP19053231314}.

If there is no plutonium on the iNavy-SEALS' world to build atomic bombs, we recommend to use ``Occam's Laser'' \citep{2007astro.ph..3783S} to blast a hole in the ice. Earthlings will use it for ''Breakthrough Starshot'' \citep{2016arXiv160401356L,2016Sci...352.1040M,2017Natur.542...20P}, a mission to $\alpha$\,Cen. A km-sized phased aperture would emit 100\,GW of laser power, sufficient to accelerate a 1\,g ``space-chip'' to $v=0.2\,$c in minutes. Light sails are no rockets, and therefore not rocket science, therefore we recommend to use the laser to blast through the ice. 

Another way to get out is to ascend through the plumes which spew from Enceladus' south polar surface. For details, see Figure~\ref{fig:ice} and Jules \citet{verne1864}.

\begin{figure}
\includegraphics[width=\linewidth]{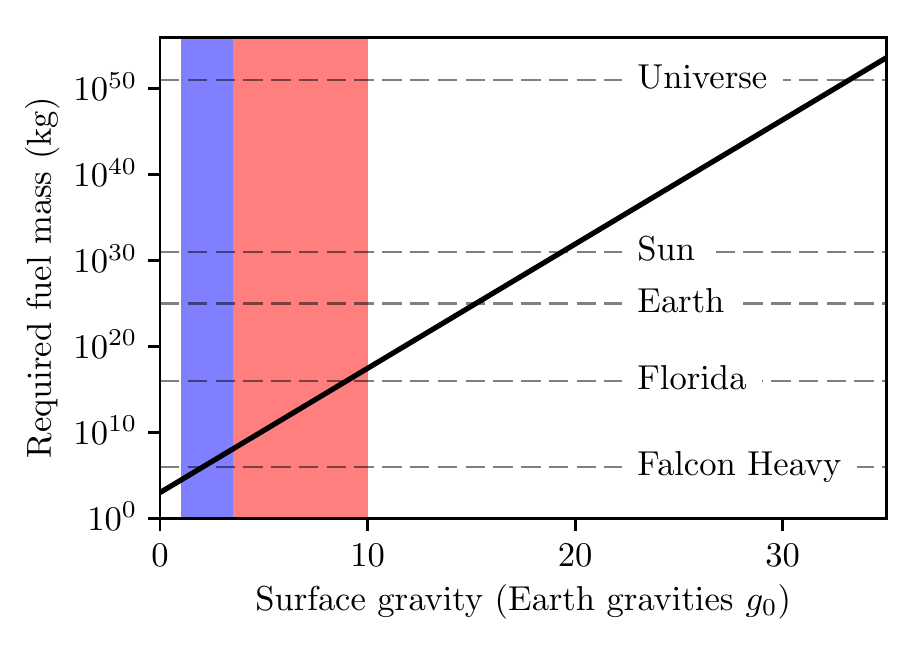}
\caption{\label{fig:gravos}Chemical fuel required for different surface gravities on Super-Earths and more heavy planets (blue and red shades). Very massive planets do not pose the question ``Will it go up?'' but ``How much of Florida go with it?''}
\end{figure}

\subsection{Amount of fuel required for different surface gravities}
For a payload of one ton to escape velocity, the required amount of chemical fuel is $\sim 3.3\,\exp(g_{\rm 0})$. The situation is not that bad for medium-sized Super-Earths, but quickly escalates due to the nasty exponential function (who likes these anyways?). On worlds with a surface gravity of $\gtrsim 10\,g_{\rm 0}$, a sizable fraction of the planet needs to be used up as chemical fuel per launch, limiting the total number of flights. We show in Figure~\ref{fig:gravos} how ridiculous the amount of fuel is for worlds with even higher surface gravity. On such worlds it is cheaper to destroy the planet rather than convert it into fuel.

In the ultimate limit, we may use the whole mass of the universe (ordinary matter only) of $\approx10^{50}\,$kg as oxygen/hydrogen fuel. Such a chemical rocket can overcome a surface gravity of $\sim 35.3\,g_{\rm 0}$. For comparison, a neutron star's density results in a very high surface gravity of $\approx10^{11}g_{\rm 0}$. Pulsar-lings will thus not become chemically space-faring beings. If such a ``universal chemical  rocket'' is launched from space directly, its final velocity would be $\sim400\,$km\,s$^{-1}$, or $\sim0.13\,$\% the speed of light. It has no trouble with interstellar dust, because its road to nowhere is free.


\section{Conclusion}
This ground-braking paper was one small step for me, but no giant leap for mankind. As such, I leave the conclusions to the reader, and suggest further research towards ice-breakers on Enceladus and Europa. I conclude with famous last words before my launch through the ice: ``May the Force be with us.''

\acknowledgments
\textit{Acknowledgments}
MH is thankful to Elon Musk and Paul Krugman for inspiration.
\end{document}